%% file: resubm.tex
\documentclass{amsart}[10pt]

\usepackage{amscd,amsmath,amssymb,latexsym,url,verbatim,algorithmic,graphicx,color}

\usepackage[dvips]{epsfig}

\title {Topology of the view complex}

\author{Dmitry N. Kozlov}

\address{Department of Mathematics, University of Bremen, 28334
  Bremen, Federal Republic of Germany}

\email{dfk@math.uni-bremen.de}

\keywords{collapses, distributed computing, combinatorial 
algebraic topology, immediate snapshot, read-write protocols}
\newtheorem{theorem}{Theorem}[section]
\newtheorem{df}[theorem]{Definition}
\newtheorem{thm}[theorem]{Theorem} 
\newtheorem{prop}[theorem]{Proposition}

 \newcommand{\nin}{\noindent}
\newcommand{\pr}{\nin{\bf Proof.} }

\newcommand{\cf}{{\mathcal F}}

\newcommand{\cs}{{\mathcal S}}
\newcommand{\csn}{\cs_{[n]}}
\newcommand{\cb}{{\mathcal B}}

\newcommand{\da}{\Delta}

\newcommand{\zz}{{\mathbb Z}}

\newcommand{\dar}{\downarrow}

\newcommand{\sm}{\setminus}
\newcommand{\view}{\textrm{\rm View}^n}
\newcommand{\is}{\chi(\da^n)}

\numberwithin{equation}{section}
\numberwithin{figure}{section}
\numberwithin{table}{section}

\begin{document}

\begin{abstract}
In this paper we consider a~family of simplicial complexes,
which we call the {\it view complexes}. Our choice of objects of study
is motivated by theoretical distributed computing, since the view
complex is a key simplicial construction used for protocol complexes
in the snapshot computational model. We show that the view complex
$\view$ can be collapsed to the well-known complex $\chi(\da^n)$,
called standard chromatic subdivision of a~simplex, and that
$\chi(\da^n)$ is itself collapsible. Furthermore, we show that the
collapses can be performed simultaneously in entire orbits of the
natural symmetric group action. Our results yield a~purely
combinatorial and constructive understanding of the topology of view
complexes, at the same time as they enhance our knowledge about 
the standard chromatic subdivision of a~simplex.
\end{abstract}

\maketitle

\section{Introduction}

Although a~close connection between theoretical distributed computing
and algebraic topology has by now been securely established, see e.g.,
\cite{AR,HS,HKR,subd,k1,K2}, many questions pertaining to arising
simplicial structures have remained unanswered. In this introduction
we make a~short excursion to the distributed context, before
proceeding with a~purely topological study in the following sections.

In theoretical distributed computing one considers $n$ processes which
are trying to solve a~task by executing a~distributed protocol.
To specify the task, one simply fixes the set of possible input
configurations, and for each input one fixes the set of legal output 
configurations. A~distributed protocol is a~sequence of instructions,
where the type of instructions, which we are allowed to use, depends
on the choice of the communication model.

Since one considers the asynchronous model, there is no such thing as 
the unique execution of the distributed protocol; instead, one has 
a~number of possible executions. One says that distributed protocol 
solves the given task, if it yields a~correct output for any input 
configuration and any possible execution. 

It has been realized, at least since~\cite{HS}, that it is fruitful 
to summarize the totality of all possible executions as a~single
simplicial complex. This complex is called the {\it protocol complex}
and depends on the distributed protocol and on the chosen communication 
model. Specifically, the protocol complex is a~pure simplicial complex,
whose dimension is one less than the number of processes. The top-dimensional
simplices of the protocol complex correspond to all possible executions 
of a~given protocol, and its vertices correspond to all possible
views of processes at the end of an execution.

Even though we focus on the shared-memory communication models, there
are still many options to choose from, some are more natural than
others. Curiously, the protocol complexes corresponding to the most
natural model, the one using write and snapshot read operations with
no restrictions on their interweaving during executions, have not been
studied much, since their simplicial structure is rather complicated.
Still, it was shown that these complexes are always contractible,
see~\cite[Chapter 10]{HKR}, \cite{Ha04}.

In contrast, the topology of the protocol complexes of the layered
immediate snapshot wait-free protocols is very easy. Namely, one can
show, that each such protocol complex is a~subdivision of a~simplex
whose vertices are indexed by the processes, see~\cite{subd,HKR}.  A
crucial construction in understanding the topology of these protocol
complexes for $n+1$ processes is the so-called {\it standard chromatic
  subdivision} $\chi(\da^n)$, see~\cite[Subsections 3.6.3, 8.4.1,
  Chapter 16]{HKR}, \cite{AR,BG,HS,subd,SZ}.

In this paper, we study the analog of this construction, which is
derived from the snapshot model. We call the corresponding simplicial
complex the {\it view complex}, see Definition~\ref{df:view} for
a~completely combinatorial description. From the point of view of
distributed computing, the view complex is a very central object,
since it is the protocol complex for the snapshot protocol in which
each process executes exactly one round.  However, for us this is just
a motivation, and we study the family of simplicial complexes
$(\view)_{n=1}^\infty$ from a purely topological point of view.

As already the lower-dimensional examples show, starting with $n=2$
the simplicial complex $\view$ does not have to be a~subdivision of
a~simplex.  As a~matter of fact, it is not a~manifold, not even
a~pseudomanifold, since its simplices of codimension $1$ may belong to
more than two top-dimensional simplices. Yet, we show that it is
possible to understand the topology of the complex $\view$ rather
completely.

To start with, it is easy to see directly that the simplicial complex
$\view$ contains $\chi(\da^n)$ as a~subcomplex. This makes sense in
the distributed computing context since every immediate snapshot
execution is also an execution in the snapshot model.  Our main
theorem, Theorem~\ref{thm:main}, states that $\view$ can be collapsed
to $\chi(\da^n)$, and that $\chi(\da^n)$ is itself collapsible. This
yields a~constructive and purely combinatorial proof of
contractibility of $\view$. However, it is stronger than the mere
contractibility, being rather a~statement about the involved
simplicial structures. We remark, that also the fact that the standard
chromatic subdvision $\chi(\da^n)$ is collapsible is new.

The simplicial complexes $\view$ and $\chi(\da^n)$ are equipped with
a~canonical simplicial action of the permutation group
$\cs_{[n]}$. This is the reflection of the fact that the considered
protocols are symmetric with respect to the renaming of
processors. The statement which we actually prove in
Theorem~\ref{thm:main} is stronger than just collapsibility. We show
that our collapses can be done in an $\cs_{[n]}$-equivariant way,
meaning that entire $\cs_{[n]}$-orbits of collapses can be performed
simultaneously, see Section~\ref{sect:coll}, and specifically
Definition~\ref{df:gcoll}, for the precise meaning of
$\cs_{[n]}$-collapses.

We recall, that in the theoretical distributed computing it is
well-known that the snapshot and the immediate snapshot models are
computationally equivalent, see e.g., \cite[Chapter 14]{HKR}.

\section{Notations} 

In order to make this paper self-contained, we shall now fix notations
and define several standard notions. To start with, for an arbitrary
positive integer $n$, we let $[n]$ denote the set
$\{0,\dots,n\}$. Furthermore, we shall use the symbols $\subset$ and
$\supset$ to denote the strict set inclusion.

\begin{df}\label{df:asc}
Let $S$ be a finite set. A family of subsets $K\subseteq 2^S$ is
called an {\bf abstract simplicial complex} if 
\begin{itemize}
\item for all $x\in S$, we have $\{x\}\in K$;
\item if $A\subset B$, and $B\in K$, then $A\in K$.
\end{itemize}
\end{df}

In the context of Definition~\ref{df:asc}, the set $S$ is called the
{\it vertex set} of $K$. Each $\sigma\in K$ is called a~{\it simplex}
of $K$. The number $|\sigma|-1$ is called the {\it dimension} of
$\sigma$ and is denoted by $\dim\sigma$. For brevity, and following
the standard practice, when $K=2^S$ we shall simply call the
corresponding abstract simplicial complex a~{\it simplex}.

There are two simplicial complexes whose vertex set is an empty set,
i.e., $S=\emptyset$, namely $K=\emptyset$, which we call the {\it void
  complex}, and $K=\{\emptyset\}$, which we call the {\it empty
  complex}. The two complexes may appear similar, but this impression
is misleading, as they have different topological properties.

\begin{figure}[hbt]

  \input{p1.pstex_t}  

\caption{A~collapsible simplicial complex, and its face poset, 
with free simplices marked with solid dots.}
\label{fig:1}
\end{figure}
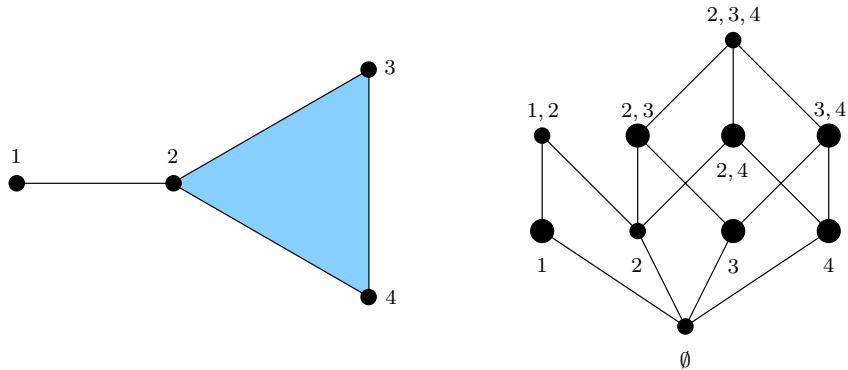

Given a simplicial complex, its simplices can be ordered by inclusion; 
the obtained partially ordered set $\cf(K)$ is called the {\it face poset} 
of $K$. For the void complex, the face poset is empty; in
all other cases the face poset has a~single minimal element, which
corresponds to the empty set. The face poset of a~simplicial complex with at
least one vertex has a~single maximal element if and only if, this
complex is a~simplex. In this case, the face poset is also called
a~{\it boolean lattice}, and is denoted by $\cb_n$, where $n$ is the
number of vertices of the simplex. Another example is shown on
Figure~\ref{fig:1}; for further details on the face poset of
a~simplicial complex we refer to~\cite[Chapter 2]{book}.

Assume now that $K$ is a~simplicial complex, and $\sigma$ is one of
its simplices. The {\it link} of $\sigma$ in $K$, is a simplicial
subcomplex of $K$ defined as follows:
\[{\textrm {lk}}_K(\sigma)=\{\tau\in K\,|\,\tau\cap\sigma=\emptyset,\,
\tau\cup\sigma\in K\}.\]

The next definition gives a~purely combinatorial description of the
notions of collapse and collapsible complex.

\begin{df}\label{df:coll}
Assume $K$ is an abstract simplicial complex and $\sigma\in K$. Set
$\cf(K)_{\geq\sigma}:=\{\tau\supseteq\sigma\,|\,\tau\in K\}$ viewed as
a~subposet of the face poset of $\cf(K)$.  We call $\sigma$ {\bf
  free}\footnote{We note that there are various notions of free simplex used
in the literature, see e.g.,~\cite{stong}.} if
$\cf(K)_{\geq\sigma}\simeq\cb_t$, for some $t\geq 1$, where $\simeq$
denotes the poset isomorphism. Alternatively, a~simplex $\sigma$ is
free if its link is a~nonempty simplex.

For a~free simplex $\sigma$, a~{\bf collapse} of $K$ associated to
$\sigma$ is the process of deleting from $K$ all the simplices in
$\cf(K)_{\geq\sigma}$.  We denote the obtained complex by
$K\dar\sigma$.  A collapse is called {\bf elementary} if
$\cf(K)_{\geq\sigma}\simeq\cb_1$.

When $M$ is a subcomplex of $K$, we say that $K$ is {\bf collapsible
  to} $M$ if there exists a sequence of collapses leading from $K$ to
$M$.  We say that $K$ is {\bf collapsible} if it is collapsible to the
void simplicial complex.
\end{df}

Figure~\ref{fig:1} illustrates the notions of free simplices and
collapsibility. Note that the void simplicial complex is collapsible,
as is any simplex, while the empty simplicial complex is not
collapsible. For a~collapsible simplicial complex the total number of
simplices in odd dimensions (including the empty simplex) must be
equal to the total number of simplices in even dimensions.

When $K$ is collapsible to $M$, we shall use the notation $K\searrow
M$. Note that in this case, there could be many different collapsing
sequences, as can be seen in the example when $K$ is a~simplex and $M$
is void.

Topologically, each collapse can be viewed as a strong deformation
retraction. In particular, if $K$ is collapsible to $M$, then one can
easily construct an explicit strong deformation retraction from 
(the geometric realization of) $K$ to~$M$. 
This means, of course, that $K$ and $M$ have the same homotopy
type, that collapsible simplicial complexes are also contractible, and
that contraction can be explicitely described.  We refer the reader
who wishes to gain deeper insight into the notion of collapse to
consult~\cite{coh} or \cite[Chapter 6]{book}.

\section{Combinatorial description of the view complex}

We now proceed defining the main objects of study of this paper.  As
mentioned in the introduction, our definition is strongly dictated by
the context of theoretical distributed computing. However, we choose
to give an~abstract description using only combinatorial topology.

\begin{df}\label{df:view}
Assume we are given a natural number $n$. An $n$-{\bf view}
is a~$2\times t$-matrix of subsets of $[n]$
\begin{equation} \label{eq:view}
\begin{pmatrix}
V_1 & \dots & V_{t-1} & [n] \\
I_1 & \dots & I_{t-1} & I_t
\end{pmatrix},
\end{equation}
where $t\geq 1$, such that the following properties are satisfied:
\begin{enumerate}
\item[(1)] $\emptyset\neq V_1\subset\dots\subset V_{t-1}\subset [n]$;
\item[(2)] the sets $I_1,\dots,I_t$ are disjoint;
\item[(3)] $\emptyset\neq I_k\subseteq V_k$, for $k=1,\dots,t-1$. 
\end{enumerate}
\end{df}

When $n$ is fixed or clear from the context, we shall simply call such
a~$2\times t$-matrix a~{\it view}. We shall also use the convention
$V_t=[n]$.

\begin{df}
Assume we have a~natural number~$n$, and an $n$-view
\[W=
\begin{pmatrix}
V_1 & \dots & V_{t-1} & [n] \\
I_1 & \dots & I_{t-1} & I_t
\end{pmatrix}.
\]
We set $\dim W:=|I_1|+\dots+|I_t|-1$, and call it the {\bf dimension}
of the view $W$.
\end{df}

Clearly, there is exactly one $n$-view of dimension $-1$, namely 
\[W=
\begin{pmatrix}
[n] \\
\emptyset
\end{pmatrix}.
\]
The $n$-views of dimension $0$ are of the form
\[
\begin{pmatrix}
\{x\}\cup A & [n] \\
\{x\}       & \emptyset 
\end{pmatrix}
\textrm{ and }
\begin{pmatrix}
 [n] \\
\{x\} 
\end{pmatrix},
\textrm{ where }\{x\}\cup A\subset[n].
\]
In any case we see that the $n$-views of dimension $0$ are indexed by
pairs $(V,x)$, where $V\subseteq [n]$ and $x\in V$. We call such
a~pair a~{\it local view}, or sometimes more specifically a~{\it local
  view of $x$}.

\begin{df}
Assume we are given an~$n$-view 
\[W=
\begin{pmatrix}
V_1 & \dots & V_{t-1} & [n] \\
I_1 & \dots & I_{t-1} & I_t
\end{pmatrix},
\]
and a~local view $L=(V,x)$. We say that $L$ {\bf belongs to} $W$,
writing $L\in W$, if there exists $1\leq k\leq t$, such that $V=V_k$
and $x\in I_k$.
\end{df}

For an arbitrary view $W$, we let $V(W)$ denote the set of all local
views belonging to $W$. Clearly, $|V(W)|=\dim W+1$.

\begin{df}\label{df:cview}
For an arbitrary natural number $n$, we define an abstract simplicial
complex $\view$ as follows:
\begin{itemize}
\item the set of vertices is the set of all local views
  \[V(\view):=\{(V,x)\,|\,x\in V\subseteq[n]\};\]
\item a subset $S\subseteq V(\view)$ forms a simplex if and only if
$S=V(W)$ for some $n$-view $W$.
\end{itemize}
\end{df}

We shall identify $n$-views with simplices of $\view$. 
\begin{prop}
The simplicial complex $\view$ is well-defined.
\end{prop}
\pr Given a~simplex $W$ of dimension $d$, one obtains all of its 
boundary simplices of dimension $d-1$ by deleting an element from one 
of the sets $I_1,\dots,I_t$. If after this the set becomes empty, one 
deletes the corresponding column in the $2\times t$-matrix, unless it 
is the last column. Clearly, what we get is again a~view, whose set 
of local views is obtained from $V(W)$ by deleting one of the
elements. Iterating this argument, we see that the conditions of
Definition~\ref{df:asc} are satisfied, and the simplicial complex
$\view$ is well-defined.
\qed 

We shall say that a~view $W$ contains a view $U$, and write $U\subseteq W$, 
if the simplex indexed by $W$ contains the simplex indexed by $U$.

\begin{figure}[hbt]

  \input{ex1b.pstex_t}  

\caption{The complexes $\view$ for $n=1$ and $n=2$.}
\label{figure:n12}
\end{figure}
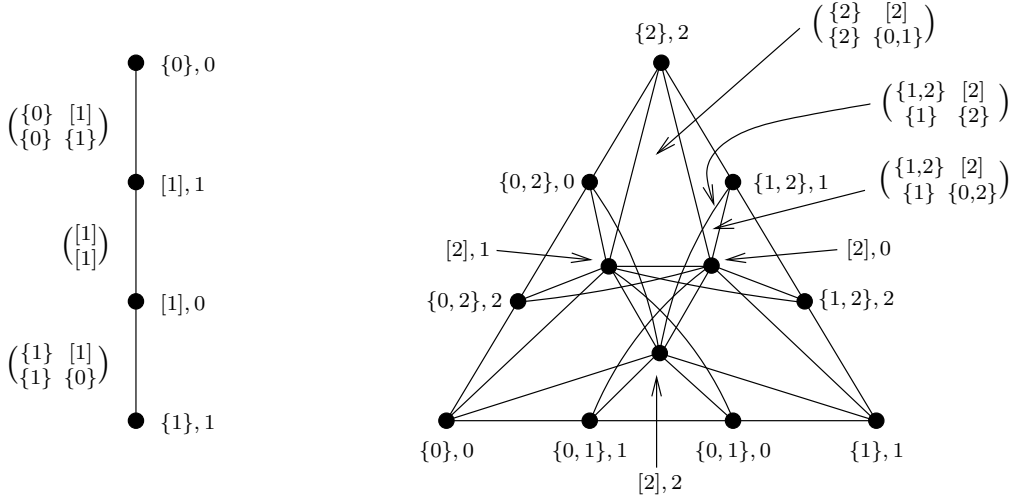

Some facts about the simplicial complex $\view$ are immediate. It is
a~{\it pure} simplicial complex of dimension~$n$, meaning that all of
its maximal simplices have dimension~$n$. It is easily seen to have
$(n+1)\cdot 2^n$ vertices. With a little more effort one can see that
$\view$ has $(n+1)\cdot n\cdot(2\cdot 3^{n-1}-2^{n-2})$ edges.  The
examples of $\view$, for $n=1$ and $n=2$, are shown on
Figure~\ref{figure:n12}.

For an~arbitrary set $A$, let $\cs_A$ denote the permutation group of
the set $A$, in particular, let $\cs_{[n]}$ denote the permutation
group of the set $[n]$. Clearly, this group is isomorphic to the
symmetric group $\cs_{n+1}$. Furthermore, there is a natural
simplicial group action of $\cs_{[n]}$ on $\view$ induced by the
permutation action on the ground set $[n]$.

We now define a~distinguished subcomplex of $\view$.

\begin{df} \label{df:is}
Assume $n$ is a natural number.

\begin{enumerate}
\item[(1)] We call an $n$-view
\[W=\begin{pmatrix}
V_1 & \dots & V_{t-1} & [n] \\
I_1 & \dots & I_{t-1} & I_t
\end{pmatrix}\] 
an {\bf immediate snapshot view} if we have $I_k\subseteq V_k\sm V_{k-1}$
for all $k=2,\dots,t$.
\item[(2)] If $W$ is an immediate snapshot view and $U\subset W$, then
  $U$ is also an immediate snapshot view. Therefore the immediate
  snapshot views form a~simplicial subcomplex of $\view$,
  which we denote by $\is$.
\end{enumerate}
\end{df}

Since the condition of being an immediate snapshot view is formulated
using the set operations only, we see that $\chi(\da^n)$ is also 
invariant under the $\cs_{[n]}$-action above.

It is furthermore clearly seen from the condition in
Definition~\ref{df:is}(1) that the difference between $\view$ and
$\is$ is first visible when $n=2$.  When $n=1$, these two complexes
are the same. When $n=2$ the complex $\view$ is obtained from $\is$ by
adding $6$ triangles of the form
\[\begin{pmatrix}
\{a,b\} & [2] \\
\{a\} & \{b,c\}
\end{pmatrix}\]
where $\{a,b,c\}=[2]$, see Figure~\ref{figure:n12}.

In general, the simplicial complex $\is$ is known as the {\it standard
  chromatic subdivision} of an $n$-simplex, see~\cite{HKR}. Its topology has been
studied in \cite{subd} where the author showed that $\is$ is a
simplicial subdivision of an $n$-simplex. The reader is invited to 
see how our description here is equivalent to the one given in 
\cite[Proposition 2.3]{subd}.

It is easy to see that the $\cs_{[n]}$-action above is the coordinate
permutation action on the standard $n$-simplex. In the context of
distributed computing it is particularly important to know that $\is$
is a~pseudomanifold. This is well-known due to the work of Herlihy and
others, see~\cite[Chapter 9]{HKR}.

On the contrary, even though the distributed computing interpretation
of the complex $\view$ is simpler than that of $\is$, understanding
its simplicial structure is harder. As a~matter of fact, it is easy to
use the nerve lemma, \cite{book}, in the same way as in \cite[Chapter
  10]{HKR}, to show that $\view$ is contractible. We do not detail
this argument here, since a~much stronger result will be shown in the
next section.  Namely, we show that $\view$ is equivariantly
collapsible, and provide an~explicit sequence of such equivariant
collapses.

  
\section{Collapsing procedure} \label{sect:coll}

As the main result of this paper, 
we shall see that $\view$ can be collapsed to $\is$. As a matter of
fact, the collapses can be done in an $\csn$-equivariant way. The
next definition formalizes this concept.

\begin{df} \label{df:gcoll}
Assume $K$ is an~abstract simplicial complex with a~simplicial action of finite
group~$G$. 
\begin{enumerate}
\item[(a)] A simplex $\sigma$ is called {\bf $G$-free} if it is free,
  and for all $g\in G$, such that $g(\sigma)\neq\sigma$, we have
\begin{equation}\label{eq:gfree}
\cf(K)_{\geq\sigma}\cap\cf(K)_{\geq g(\sigma)}=\emptyset.
\end{equation}
\item[(b)] If $\sigma$ is $G$-free, we call the procedure of deleting
  all the simplices from the union $\bigcup_{g\in G}\cf(K)_{\geq
    g(\sigma)}$ the {\bf $G$-collapse} of~$K$.
\end{enumerate}
\end{df}

Note, that when $\sigma$ is free, each $g(\sigma)$ is automatically
free as well. Therefore, deleting the simplices from $\cf(K)_{\geq
  g(\sigma)}$ is also a~collapse. The condition~\eqref{eq:gfree}
guarantees that all these collapses can be done simultaneously and
independently of each other. This is because for all $g,h\in G$, 
whenever $\tau$ is a~simplex of $K$, the identity 
\[\cf(K)_{\geq h_1(\tau)}\cap\cf(K)_{\geq h_2(\tau)}=\emptyset\]
follows from~\eqref{eq:gfree} by substituting $\sigma:=h_1(\tau)$ 
and $g:=h_2\circ h_1^{-1}$. In particular, we see that geometrically
a~$G$-collapse yields a~$G$-equivariant strong deformation retraction.

\begin{figure}[hbt]

  \input{ex1d.pstex_t}  

\caption{Two simplicial complexes with $\zz_2$-action. The end points 
are $\zz_2$-free in the second, but not in the first complex.}
\label{fig:act}
\end{figure}
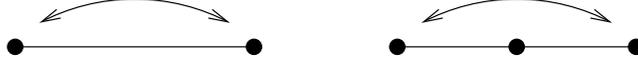

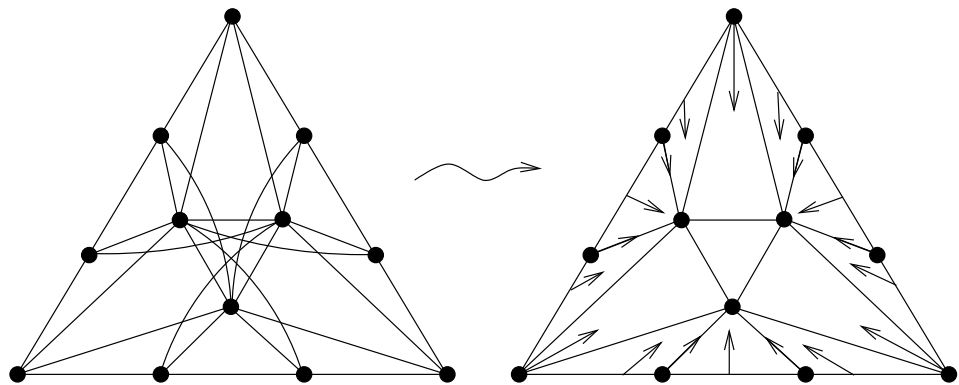
\begin{figure}[hbt]

  \input{ex1c.pstex_t}  

\caption{The collapsing procedure for ${\textrm{\rm View}^2}$. The first step collapses
the 6 extraordinary triangles.}
\label{fig:ex1c}
\end{figure}

The simplest example of a~free simplex which is not $G$-free is given
by taking $K$ to be a~$1$-simplex, and letting $G=\zz_2$ act on $K$ by
swapping the vertices, see Figure~\ref{fig:act}. Each vertex is free,
and leads to an (elementary) collapse, but they are not $\zz_2$-free,
and the collapses cannot be performed simultaneously. As shown on the
same figure, subdividing the interval in the middle leads to
a~$\zz_2$-complex, in which both end vertices are $\zz_2$-free. A~more
complicated example of $\cs_3$-collapsing sequence is shown on
Figure~\ref{fig:ex1c}.

In analogy with the usual collapses we shall say that $K$ is {\it
  $G$-collapsible} to a~subcomplex $M$ is there exists a sequence of
$G$-collapses leading from $K$ to $M$. In this case $M$ must be
$G$-invariant, and we use the notation $K\searrow_G M$. Same way, we
say that $K$ is $G$-collapsible if it is $G$-collapsible to a void
complex.

An example of an $\csn$-collapsible complex is the $n$-simplex
$\da^n$.  However, in contrast to the situation with regular
collapses, there is a~unique $\csn$-collapsing sequence from $\da^n$
to the void complex. Namely, we must collapse $\da^n$ to the void
complex in one single step.

We now proceed to define two functions on the set of views, which will
be crucial for constructing our collapsing sequence.

\begin{df}\label{df:phi}
For an arbitrary $n$-view
\begin{equation}\label{eq:w}
W=
\begin{pmatrix}
V_1 & \dots & V_{t-1} & [n] \\
I_1 & \dots & I_{t-1} & I_t
\end{pmatrix},
\end{equation}
we set 
\[\Phi(W):=
\begin{pmatrix}
V_1 & \dots & V_{t-1} & [n] \\ 
I_1 & \dots & I_{t-1} & I_t\cap V_{t-1}
\end{pmatrix},
\]
and
\[\Psi(W):=
\begin{pmatrix}
V_1 & \dots & V_{t-1} & [n] \\ 
I_1 & \dots & I_{t-1} & I_t\cup([n]\sm V_{t-1})
\end{pmatrix}.
\]
Furthermore, we let $I(W)$ denote the closed interval
$[\Phi(W),\Psi(W)]$ in the face poset of $\view$.
\end{df}

\begin{figure}[hbt]

  \input{phi.pstex_t}  

\caption{The functions $\Phi$ and $\Psi$.}
\label{fig:phi}
\end{figure}
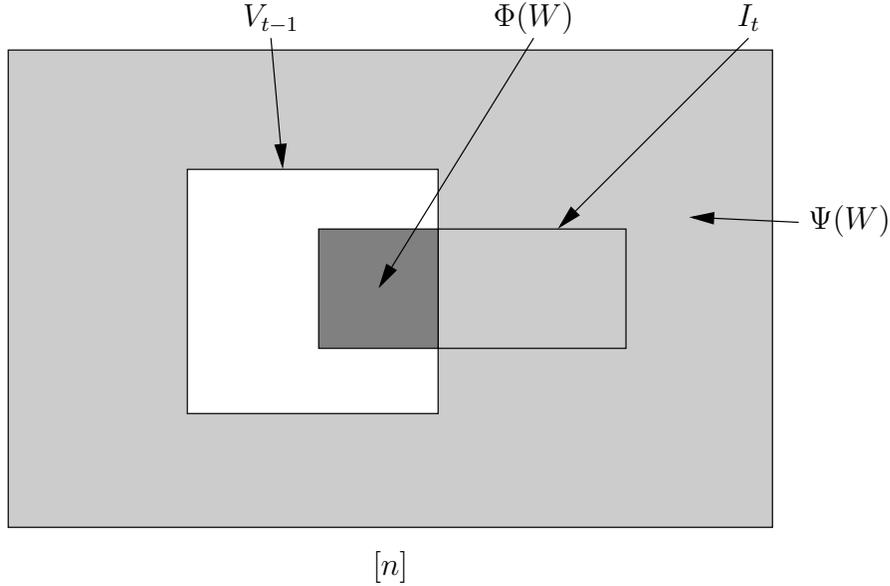

In particular, we allow $\Phi(W)$ to be the empty simplex. This is the
case when $I_1=\dots=I_{t-1}=I_t\cap V_{t-1}=\emptyset$, which is
equivalent to saying that $t=1$, in other words $W=\begin{pmatrix} [n]
\\ I \end{pmatrix}$, for some $I\subseteq[n]$. The
Definition~\ref{df:phi} is illustrated on Figure~\ref{fig:phi}.

\begin{prop} \label{prop:phi}
The maps $\Phi$ and $\Psi$ enjoy the following properties
\begin{enumerate}
\item[(1)] $\Phi(W)\subseteq W\subseteq\Psi(W)$, for all views $W$;
\item[(2)] $\Phi(W)\subseteq U\subseteq\Psi(W)$ implies
  $\Phi(U)=\Phi(W)$ and $\Psi(U)=\Psi(W)$;
\item[(3)] $\Phi(W)\neq\Psi(W)$, for all views $W$;  
\item[(4)] if $U\subseteq W$, then $\Phi(U)\subseteq\Phi(W)$;
\item[(5)] for all views $W$ we have the following implication: 
if $\Phi(W)\in\cf(\chi(\da^n))$ then $W\in\cf(\chi(\da^n))$;
\item[(6)] for all $W\in\view$, and all $\pi\in\cs_{[n]}$, we have $\pi(I(W))=I(\pi(W))$.
\end{enumerate}
\end{prop}
\pr To see (1), note that the view $\Phi(W)$ is obtained from $W$ by
removing the local views of the form $([n],x)$, such that $x\in I_t\sm
V_{t-1}$, hence $\Phi(W)\subseteq W$. The view $\Psi(W)$ is obtained
from $W$ by adding local views of the form $([n],x)$, such that
$x\notin I_t\cup V_{t-1}$, hence $\Psi(W)\supseteq W$.

Let us now show (2). Assume $\Phi(W)\subseteq U\subseteq\Psi(W)$,
where $W$ is a~view given in~\eqref{eq:w}. The difference between
$\Phi(W)$ and $\Psi(W)$ consists if all local views $([n],x)$, such
that $x\notin V_{t-1}$, see Figure~\ref{fig:phi}. This means, that
there exists a~set $S$ satisfying $S\cap V_{t-1}=\emptyset$, such that
\[U=
\begin{pmatrix}
V_1 & \dots & V_{t-1} & [n] \\
I_1 & \dots & I_{t-1} & (I_t\cap V_{t-1})\cup S
\end{pmatrix}.\]
Clearly, 
\[V_{t-1}\cap((I_t\cap V_{t-1})\cup S)=(I_t\cap
V_{t-1})\cup(V_{t-1}\cap S)=I_t\cap V_{t-1},\] 
so $\Phi(U)=\Phi(W)$. Furthermore, 
\[(I_t\cap V_{t-1})\cup S\cup([n]\sm V_{t-1})=(I_t\cap
V_{t-1})\cup([n]\sm V_{t-1}),\] 
so $\Psi(U)=\Psi(W)$.

To see (3) we note that we always have $V_{t-1}\neq[n]$, hence there
exists at least one local view $([n],x)\in V(W)$, such that $x\notin
V_{t-1}$.

To show (4) assume $U\subseteq W$. If $U$ is the empty simplex, then
the statement is obvious, so let us assume $U$ is not the empty
simplex. Consider the presentations
\[W=\begin{pmatrix}
V_1 & \dots & V_{t-1} & [n] \\
I_1 & \dots & I_{t-1} & I_t
\end{pmatrix}, \quad
U=\begin{pmatrix}
V_1 & \dots & V_{t-1} & [n] \\
J_1 & \dots & J_{t-1} & J_t
\end{pmatrix},\]
where $J_k\subseteq I_k$, for all $k=1,\dots,t$. For the view $W$ the
presentation above is standard, but not necessarily for $U$. Namely,
we allow some of $J_k$'s to be empty. If they are, we can simply
delete the corresponding column (unless it is the last column) to
obtain the standard presentation on an $n$-view that we have used so
far.

Let $1\leq k\leq t-1$ be the largest index less than $t$, for which
$J_k\neq\emptyset$, and set $S:=V_k$. If
$J_1=\dots=J_{t-1}=\emptyset$, we set $S:=\emptyset$.  We have
\[\Phi(U)=
\begin{pmatrix}
V_1 & \dots & V_{t-1} & [n] \\
J_1 & \dots & J_{t-1} & J_t\cap S
\end{pmatrix},\]
where again we use the presentation where empty sets in the second row
are allowed. In any case, we have $J_t\subseteq I_t$, and $S\subseteq
V_{t-1}$, hence $J_t\cap S\subseteq I_t\cap V_{t-1}$. Together with
the fact that $J_k\subseteq I_k$, for $k=1,\dots,t-1$ this implies
$\Phi(U)\subseteq\Phi(W)$.

To see (5) note that, since $W$ differs from $\Phi(W)$ only in $I_t$,
we only need to check the condition from Definition~\ref{df:is}(1) for
this set. As noticed earlier, when going from $\Phi(W)$ to $W$, the
set $I_t$ is enlarged by some elements $x\notin V_{t-1}$. Hence the
condition $I_t\subseteq[n]\sm V_{t-1}$ will be satisfied for $W$ as
well, and we may conclude that $W$ is an immediate snapshot view, if
$\Phi(W)$ is one.

To see (6) note that the definitions of $\Phi(W)$ and $\Psi(W)$, hence also
the definition of the interval $I(W)$, is formulated completely in terms of 
set operations, see also Figure~\ref{fig:phi}. The $\cs_{[n]}$-action is 
simply renaming of the elements of the underlying set $[n]$, so 
$\pi(I(W))=I(\pi(W))$.
\qed
\vspace{5pt}

Note that Proposition~\ref{prop:phi}(6) implies that in particular 
$\Phi(\pi(W))=\pi(\Phi(W))$ and $\Psi(\pi(W))=\pi(\Psi(W))$. Furthermore, 
since the intervals $\pi(I(W))$ and $I(W)$ are either equal or disjoint, 
we see that if $\Phi(W)\neq\pi(\Phi(W))$, then $I(W)$ is disjoint from 
$I(\pi(W))$.

We are now ready to prove our main theorem.

\begin{thm} \label{thm:main}
For every natural number $n$, the following statements are true.
\begin{enumerate}
\item[(1)] The simplicial complex $\view$ is $\csn$-collapsible to
  $\is$.
\item[(2)] The simplicial complex $\is$ is $\csn$-collapsible.
\end{enumerate}
\end{thm}
\pr Proposition~\ref{prop:phi}(1) implies that
$\Phi(W)\subseteq\Psi(W)$, for all views $W$.  Note, that as a~poset
$I(W)\simeq\cb_{|\Psi(W)|-|\Phi(W)|}$.  Since $W\in I(W)$, for all
views $W$, the intervals $I(W)$ cover the poset $\cf(\view)$.

On the other hand, Proposition~\ref{prop:phi}(2) shows that either
$I(W)=I(U)$, or $I(W)\cap I(U)=\emptyset$, for all views $U,W$. So
$\cf(\view)$ can, in fact, be decomposed as a~disjoint union of
intervals $I(W_1),\dots,I(W_c)$, for some choice of views
$W_1,\dots,W_c$. Without loss of generality we can assume that
$\Phi(W_1)=W_1$, $\dots$, $\Phi(W_c)=W_c$, and
$|W_1|\geq|W_2|\geq\dots\geq|W_c|$. By Proposition~\ref{prop:phi}(2)
we see that for all $U\in\cf(\view)$ there exists $k$ between $1$ and
$c$ such that $\Phi(U)=W_k$. We now want to show that starting with
$\view$, collapsing first $W_1$, then $W_2$, and so on, until $W_c$,
will yield a~collapsing sequence from $\view$ to the void complex.

For every $0\leq k\leq c$ let $V_k$ be the subcomplex of $\view$
consisting of all simplices $\tau$, such that $\tau\not\supseteq W_i$,
for all $i=1,\dots,k$. In particular, $V_0=\view$.  We shall show by
induction on~$k$, $1\leq k\leq c$, that
\begin{equation}
\label{eq:star1}
\cf(V_{k-1})_{\geq W_K}=I(W_k).
\end{equation}  
We start with $k=1$. If $U\supseteq W_1$, then
Proposition~\ref{prop:phi}(4) implies that
$\Phi(U)\supseteq\Phi(W_1)=W_1$. But we know that $\Phi(U)=W_l$, for
some $l$, hence we get $l=1$, and $\Phi(U)=W_1$. This means that $U\in
I(W_1)$. Altogether this implies that $\cf(\view)_{\geq W_1}=I(W_1)$.

For an induction step, let $2\leq k\leq c$. Assume $U\supseteq W_k$,
then as before $\Phi(U)\supseteq\Phi(W_k)=W_k$. Pick $l$ such that
$\Phi(U)=W_l$. If $k=l$, then $\Phi(U)=W_k$, hence $U\in
I(W_k)$. Else, we must have $W_l\supset W_k$, in particular
$|W_l|>|W_k|$. This implies $k>l$, and so $U\notin V_{k-1}$. This
shows that $\cf(V_{k-1})_{\geq W_k}\subseteq I(W_k)$. The other
direction follows from the fact that the intervals are disjoint,
together with the induction hypothesis. Namely, we have
$\cf(V_{k-1})=\cf(\view)\sm\cup_{i=1}^{k-1}I(W_i)$, which implies
$I(W_k)\subseteq\cf(V_{k-1})$, and hence
$I(W_k)\subseteq\cf(V_{k-1})_{\geq W_k}$. Summarizing, we conclude
that \eqref{eq:star1} holds for this~$k$.

The equality \eqref{eq:star1} means that for every $1\leq k\leq c$,
the simplex $W_k$ is free in $V_{k-1}$, and that the corresponding
collapse results in $V_k$.  Proposition~\ref{prop:phi}(3) says that
$\Phi(W)\neq\Psi(W)$, in particular $I(W)\simeq\cb_t$, for $t\geq 1$,
so the condition of Definition~\ref{df:coll} is satisfied.

Since we already saw that $\cf(\view)$ is a~disjoint union of the
intervals $I(W_1),\dots,I(W_c)$, we conclude that $\view$ is
collapsible.  

We shall now adjust the collapsing sequence above to first lead to
$\chi(\da^n)$, and then collapse $\chi(\da^n)$. First, note that the
condition $|W_1|\geq\dots\geq|W_c|$ was strictly speaking
unnecessarily strong to be able to conclude that we have a~collapsing
sequence. All we needed was the implication that if $W_l\supset W_k$,
then $l<k$, so any linear extension of the set $\{W_1,\dots,W_c\}$,
equipped with the reverse inclusion order, would do. On the other
hand, note that Proposition~\ref{prop:phi}(5) implies that for all $W$
either $I(W)\subseteq\cf(\chi(\da^n))$, or $I(W)$ and
$\cf(\chi(\da^n))$ are disjoint.

Since $\chi(\da^n)$ is a~simplicial subcomplex of $\view$, its face
poset is a~lower ideal of $\cf(\view)$. In particular, the linear
extension of the set $\{W_1,\dots,W_c\}$ can be chosen in a~special
way: first take any linear extension of the subset
$\{W_1,\dots,W_c\}\cap(\cf(\view)\sm\cf(\chi(\da^n)))$, and then
concatenate it with any linear extension of
$\{W_1,\dots,W_c\}\cap\cf(\chi(\da^n))$. By what was said above, this
concatenation is a~linear extension by itself. This linear extension
now yields a~collapsing sequence from $\view$ to $\chi(\da^n)$, and
then from $\chi(\da^n)$ to the void complex.

To finish the proof of our theorem, we need to modufy the collapsing
sequence once more, in order to incorporate the $\cs_{[n]}$-action.
Until now we did everything for an arbitrary set $\{W_1,\dots,W_c\}$
such that the intervals $I(W_1),\dots,I(W_c)$ cover $\cf(\view)$. We
shall now specify this set.

Set $L:=\{\Phi(W)\,|\,W\in\view,\,\,W\notin\chi(\da^n)\}$, and pick
$W_1$ such that $|W_1|=\max_{W\in L}|W|$. By what is said above, $W_1$
is free in $\view$. Let $\{W_1,\dots,W_p\}$ be any set of
representatives of the orbit $\cs_{[n]}(W_1)$. Specifically, this
means that the views $W_1,\dots,W_p$ are all distinct, and for each
$\pi\in\cs_{[n]}$ there exists $i$, $1\leq i\leq p$, such that
$\pi(W_1)=W_i$. By Proposition~\ref{prop:phi}(6), we know that the
intervals $I(W_1),\dots,I(W_p)$ are disjoint, and for each
$\pi\in\cs_{[n]}$ there exists $i$, $1\leq i\leq p$, such that
$\pi(I(W_1))=I(W_i)$. In other words, $I(W_1),\dots,I(W_p)$ is a~set
of representatives of the orbit $\cs_{[n]}(I(W_1))$.

We can now repeat the entire procedure with $\cf(\view)\sm\cup_{i=1}^p
I(W_i)$, and then proceed iterating until the entire difference
$\cf(\view)\sm\cf(\chi(\da^n))$ is covered with chosen
intervals. After that we proceed to do the same for~$\chi(\da^n)$.

As a~final result, we will obtain a~sequence of collapses such that
\begin{itemize}
\item the collapses come in $\cs_{[n]}$-equivariant batches; in each
  such batch, all collapses can be performed simultaneously, resulting
  in an~$\cs_{[n]}$-collapse;
\item the collapses first exhaust the difference
  $\cf(\view)\sm\cf(\chi(\da^n))$, and then proceed to collapse the
  simplicial complex $\chi(\da^n)$.
\end{itemize}
The arguments above adopt easily to this specific sequence of
collapses, so we are finished with the proof of the theorem. 
\qed

\vskip5pt

\nin For $n=2$ we illustrate the collapsing procedure from
Theorem~\ref{thm:main} on Figure~\ref{fig:ex1c}.

\end{document}

%% file: p1.pstex_t
\begin{picture}(0,0)%
\includegraphics{p1.pstex}%
\end{picture}%
\setlength{\unitlength}{3947sp}%
\begingroup\makeatletter\ifx\SetFigFont\undefined%
\gdef\SetFigFont#1#2#3#4#5{%
  \reset@font\fontsize{#1}{#2pt}%
  \fontfamily{#3}\fontseries{#4}\fontshape{#5}%
  \selectfont}%
\fi\endgroup%
\begin{picture}(5236,2325)(-205,-1567)
\put(831,-236){\makebox(0,0)[b]{\smash{{\SetFigFont{8}{9.6}{\rmdefault}{\mddefault}{\updefault}{\color[rgb]{0,0,0}$2$}%
}}}}
\put(-151,-232){\makebox(0,0)[b]{\smash{{\SetFigFont{8}{9.6}{\rmdefault}{\mddefault}{\updefault}{\color[rgb]{0,0,0}$1$}%
}}}}
\put(2166,-1121){\makebox(0,0)[lb]{\smash{{\SetFigFont{8}{9.6}{\rmdefault}{\mddefault}{\updefault}{\color[rgb]{0,0,0}$4$}%
}}}}
\put(2161,324){\makebox(0,0)[lb]{\smash{{\SetFigFont{8}{9.6}{\rmdefault}{\mddefault}{\updefault}{\color[rgb]{0,0,0}$3$}%
}}}}
\put(3151,-916){\makebox(0,0)[b]{\smash{{\SetFigFont{8}{9.6}{\rmdefault}{\mddefault}{\updefault}{\color[rgb]{0,0,0}$1$}%
}}}}
\put(3746,-916){\makebox(0,0)[b]{\smash{{\SetFigFont{8}{9.6}{\rmdefault}{\mddefault}{\updefault}{\color[rgb]{0,0,0}$2$}%
}}}}
\put(4051,-1516){\makebox(0,0)[b]{\smash{{\SetFigFont{8}{9.6}{\rmdefault}{\mddefault}{\updefault}{\color[rgb]{0,0,0}$\emptyset$}%
}}}}
\put(4351,-921){\makebox(0,0)[b]{\smash{{\SetFigFont{8}{9.6}{\rmdefault}{\mddefault}{\updefault}{\color[rgb]{0,0,0}$3$}%
}}}}
\put(4951,-916){\makebox(0,0)[b]{\smash{{\SetFigFont{8}{9.6}{\rmdefault}{\mddefault}{\updefault}{\color[rgb]{0,0,0}$4$}%
}}}}
\put(3156, 59){\makebox(0,0)[b]{\smash{{\SetFigFont{8}{9.6}{\rmdefault}{\mddefault}{\updefault}{\color[rgb]{0,0,0}$1,2$}%
}}}}
\put(3751, 54){\makebox(0,0)[b]{\smash{{\SetFigFont{8}{9.6}{\rmdefault}{\mddefault}{\updefault}{\color[rgb]{0,0,0}$2,3$}%
}}}}
\put(4961, 64){\makebox(0,0)[b]{\smash{{\SetFigFont{8}{9.6}{\rmdefault}{\mddefault}{\updefault}{\color[rgb]{0,0,0}$3,4$}%
}}}}
\put(4346,-321){\makebox(0,0)[b]{\smash{{\SetFigFont{8}{9.6}{\rmdefault}{\mddefault}{\updefault}{\color[rgb]{0,0,0}$2,4$}%
}}}}
\put(4356,659){\makebox(0,0)[b]{\smash{{\SetFigFont{8}{9.6}{\rmdefault}{\mddefault}{\updefault}{\color[rgb]{0,0,0}$2,3,4$}%
}}}}
\end{picture}%

%% file: ex1b.pstex_t
\begin{picture}(0,0)%
\includegraphics{ex1b.pstex}%
\end{picture}%
\setlength{\unitlength}{3947sp}%
\begingroup\makeatletter\ifx\SetFigFont\undefined%
\gdef\SetFigFont#1#2#3#4#5{%
  \reset@font\fontsize{#1}{#2pt}%
  \fontfamily{#3}\fontseries{#4}\fontshape{#5}%
  \selectfont}%
\fi\endgroup%
\begin{picture}(4871,3084)(1486,-2342)
\put(3601,-2086){\makebox(0,0)[b]{\smash{{\SetFigFont{8}{9.6}{\rmdefault}{\mddefault}{\updefault}{\color[rgb]{0,0,0}$\{0\},0$}%
}}}}
\put(4501,-2086){\makebox(0,0)[b]{\smash{{\SetFigFont{8}{9.6}{\rmdefault}{\mddefault}{\updefault}{\color[rgb]{0,0,0}$\{0,1\},1$}%
}}}}
\put(5401,-2086){\makebox(0,0)[b]{\smash{{\SetFigFont{8}{9.6}{\rmdefault}{\mddefault}{\updefault}{\color[rgb]{0,0,0}$\{0,1\},0$}%
}}}}
\put(6301,-2086){\makebox(0,0)[b]{\smash{{\SetFigFont{8}{9.6}{\rmdefault}{\mddefault}{\updefault}{\color[rgb]{0,0,0}$\{1\},1$}%
}}}}
\put(4951,539){\makebox(0,0)[b]{\smash{{\SetFigFont{8}{9.6}{\rmdefault}{\mddefault}{\updefault}{\color[rgb]{0,0,0}$\{2\},2$}%
}}}}
\put(3951,-1146){\makebox(0,0)[rb]{\smash{{\SetFigFont{8}{9.6}{\rmdefault}{\mddefault}{\updefault}{\color[rgb]{0,0,0}$\{0,2\},2$}%
}}}}
\put(4406,-396){\makebox(0,0)[rb]{\smash{{\SetFigFont{8}{9.6}{\rmdefault}{\mddefault}{\updefault}{\color[rgb]{0,0,0}$\{0,2\},0$}%
}}}}
\put(5526,-401){\makebox(0,0)[lb]{\smash{{\SetFigFont{8}{9.6}{\rmdefault}{\mddefault}{\updefault}{\color[rgb]{0,0,0}$\{1,2\},1$}%
}}}}
\put(5936,-1131){\makebox(0,0)[lb]{\smash{{\SetFigFont{8}{9.6}{\rmdefault}{\mddefault}{\updefault}{\color[rgb]{0,0,0}$\{1,2\},2$}%
}}}}
\put(4941,-2291){\makebox(0,0)[b]{\smash{{\SetFigFont{8}{9.6}{\rmdefault}{\mddefault}{\updefault}{\color[rgb]{0,0,0}$[2],2$}%
}}}}
\put(6111,-816){\makebox(0,0)[lb]{\smash{{\SetFigFont{8}{9.6}{\rmdefault}{\mddefault}{\updefault}{\color[rgb]{0,0,0}$[2],0$}%
}}}}
\put(3876,-821){\makebox(0,0)[rb]{\smash{{\SetFigFont{8}{9.6}{\rmdefault}{\mddefault}{\updefault}{\color[rgb]{0,0,0}$[2],1$}%
}}}}
\put(1804,338){\makebox(0,0)[lb]{\smash{{\SetFigFont{8}{9.6}{\rmdefault}{\mddefault}{\updefault}{\color[rgb]{0,0,0}$\{0\},0$}%
}}}}
\put(1804,-415){\makebox(0,0)[lb]{\smash{{\SetFigFont{8}{9.6}{\rmdefault}{\mddefault}{\updefault}{\color[rgb]{0,0,0}$[1],1$}%
}}}}
\put(1804,-1162){\makebox(0,0)[lb]{\smash{{\SetFigFont{8}{9.6}{\rmdefault}{\mddefault}{\updefault}{\color[rgb]{0,0,0}$[1],0$}%
}}}}
\put(1804,-1912){\makebox(0,0)[lb]{\smash{{\SetFigFont{8}{9.6}{\rmdefault}{\mddefault}{\updefault}{\color[rgb]{0,0,0}$\{1\},1$}%
}}}}
\put(1501,-811){\makebox(0,0)[rb]{\smash{{\SetFigFont{12}{14.4}{\rmdefault}{\mddefault}{\updefault}{\color[rgb]{0,0,0}${[1]}\choose{[1]}$}%
}}}}
\put(1501,-61){\makebox(0,0)[rb]{\smash{{\SetFigFont{12}{14.4}{\rmdefault}{\mddefault}{\updefault}{\color[rgb]{0,0,0}${\{0\}\,\,\,\,[1]\,}\choose{\{0\}\,\,\,\{1\}}$}%
}}}}
\put(1501,-1561){\makebox(0,0)[rb]{\smash{{\SetFigFont{12}{14.4}{\rmdefault}{\mddefault}{\updefault}{\color[rgb]{0,0,0}${\{1\}\,\,\,\,[1]\,}\choose{\{1\}\,\,\,\{0\}}$}%
}}}}
\put(6301,-395){\makebox(0,0)[lb]{\smash{{\SetFigFont{12}{14.4}{\rmdefault}{\mddefault}{\updefault}{\color[rgb]{0,0,0}${\{1,2\}\,\,\,\,[2]\,\,\,}\choose{\,\,\,\{1\}\,\,\,\{0,2\}}$}%
}}}}
\put(6316, 88){\makebox(0,0)[lb]{\smash{{\SetFigFont{12}{14.4}{\rmdefault}{\mddefault}{\updefault}{\color[rgb]{0,0,0}${\{1,2\}\,\,\,\,[2]\,\,\,}\choose{\,\,\,\{1\}\,\,\,\,\,\{2\}\,\,}$}%
}}}}
\put(5881,583){\makebox(0,0)[lb]{\smash{{\SetFigFont{12}{14.4}{\rmdefault}{\mddefault}{\updefault}{\color[rgb]{0,0,0}${\{2\}\,\,\,\,\,\,[2]\,\,\,}\choose{\,\{2\}\,\,\,\{0,1\}}$}%
}}}}
\end{picture}%

%% file: ex1d.pstex_t
\begin{picture}(0,0)%
\includegraphics{ex1d.pstex}%
\end{picture}%
\setlength{\unitlength}{3947sp}%
\begingroup\makeatletter\ifx\SetFigFont\undefined%
\gdef\SetFigFont#1#2#3#4#5{%
  \reset@font\fontsize{#1}{#2pt}%
  \fontfamily{#3}\fontseries{#4}\fontshape{#5}%
  \selectfont}%
\fi\endgroup%
\begin{picture}(4012,364)(1595,-1916)
\end{picture}%

%% file: phi.pstex_t
\begin{picture}(0,0)%
\includegraphics{phi.pstex}%
\end{picture}%
\setlength{\unitlength}{3947sp}%
\begingroup\makeatletter\ifx\SetFigFont\undefined%
\gdef\SetFigFont#1#2#3#4#5{%
  \reset@font\fontsize{#1}{#2pt}%
  \fontfamily{#3}\fontseries{#4}\fontshape{#5}%
  \selectfont}%
\fi\endgroup%
\begin{picture}(5052,3678)(1789,-2980)
\put(5101,539){\makebox(0,0)[b]{\smash{{\SetFigFont{12}{14.4}{\rmdefault}{\mddefault}{\updefault}{\color[rgb]{0,0,0}$\Phi(W)$}%
}}}}
\put(4201,-2911){\makebox(0,0)[b]{\smash{{\SetFigFont{12}{14.4}{\rmdefault}{\mddefault}{\updefault}{\color[rgb]{0,0,0}$[n]$}%
}}}}
\put(6826,-736){\makebox(0,0)[lb]{\smash{{\SetFigFont{12}{14.4}{\rmdefault}{\mddefault}{\updefault}{\color[rgb]{0,0,0}$\Psi(W)$}%
}}}}
\put(3451,539){\makebox(0,0)[b]{\smash{{\SetFigFont{12}{14.4}{\rmdefault}{\mddefault}{\updefault}{\color[rgb]{0,0,0}$V_{t-1}$}%
}}}}
\put(6451,539){\makebox(0,0)[b]{\smash{{\SetFigFont{12}{14.4}{\rmdefault}{\mddefault}{\updefault}{\color[rgb]{0,0,0}$I_t$}%
}}}}
\end{picture}%

%% file: ex1c.pstex_t
\begin{picture}(0,0)%
\includegraphics{ex1c.pstex}%
\end{picture}%
\setlength{\unitlength}{3947sp}%
\begingroup\makeatletter\ifx\SetFigFont\undefined%
\gdef\SetFigFont#1#2#3#4#5{%
  \reset@font\fontsize{#1}{#2pt}%
  \fontfamily{#3}\fontseries{#4}\fontshape{#5}%
  \selectfont}%
\fi\endgroup%
\begin{picture}(5962,2360)(395,-1916)
\end{picture}%